\documentclass{ptapap}

\author{Bogumi{\l} Pilecki}[CAMK]
\affil[CAMK]{Nicolaus Copernicus Astronomical Center\\
  Bartycka 18, 00--716 Warszawa, Poland}

\title{Cepheids in Eclipsing Binaries. What and How We Can Learn About Them}
\headtitle{Cepheids in Eclipsing Binaries}
  
\begin{document}

\maketitle

\begin{abstract}
Eclipsing binary systems with pulsating components offer a unique possibility to accurately measure the most important parameters of pulsating stars, to study their evolution, and to test the pulsation theory. I will show what we can learn about the pulsating stars from the analysis of such systems and how we can do it. Special attention will be paid to the mass, radius, $p$-factor, and distance determination. Although the core of the method is based on the observations of double-lined eclipsing spectroscopic binaries, with the help of the pulsation theory, it is possible to measure absolute parameters for single-lined binaries also.
\end{abstract}

\section{Introduction}\label{sec:intro}

About 60\% of Galactic Cepheids are suspected to be members of binary systems \citep{Evans_2015_Cepheid_Binarity_AJ.150.13E}. For them, the light of the companion adds to the observed Cepheid brightness, which may introduce systematic errors in the period-luminosity relations. It is also a problem for the parallax measurements (e.g.\ by the {\it Gaia} mission) -- the orbital motion has to be taken into account to ensure proper distance determination.

The biggest issue is when the companion cannot be observed (e.g.\ Cepheids in non-eclipsing, single-lined binary systems), and only the negative effects of binarity are present. On the other hand, if a Cepheid in a detached, eclipsing, double-lined spectroscopic binary is observed, very precise physical parameters (like mass, radius, and luminosity) can be determined for the system components. Moreover, a distance measurement using almost purely geometrical methods is possible.

Unfortunately, for a long time no Cepheid was found in an eclipsing binary system. The breakthrough came from the microlensing surveys, MACHO \citep{macho2002alcock} and OGLE \citep{ogle2015udalski}. More than ten candidates in the Small (SMC) and Large (LMC) Magellanic Clouds were proposed after the analysis of their light curves; however, spectroscopic confirmations were needed to remove all doubt.  We have selected the most promising objects and observed them spectroscopically.

OGLE LMC-CEP-0227 was the first to be confirmed and analyzed \citep{cep227nature2010}, followed by six other Cepheids for which the parameters were measured: OGLE LMC-CEP-1812 \citep{cep1812apj2011}, two Cepheids in OGLE LMC-CEP-1718 \citep{cep1718apj2014}, OGLE LMC-CEP-2532 \citep{cep2532apj2015}, OGLE LMC562.05.9009 \citep{cep9009apj2015} and OGLE LMC-T2CEP-098 \citep{t2cep098apj2017}. OGLE LMC-CEP-0227 was later reanalyzed by \citet{cep227mnras2013} using more data and a more advanced method.

\section{Analysis}

As mentioned above, for a Cepheid in a binary system the observed flux is contaminated with the light of the companion, and there is an additional orbital motion. The variability is more complex, which affects both light and radial velocity curves. These factors make the analysis much harder than for a normal eclipsing binary system, but compared with the analysis of a single Cepheid, it gives us a lot more information.

The classification of the Cepheid can be done regardless of its binarity as it depends mostly on the shape of the light curve for a given period and not as much on the amplitude. Nevertheless, the subtraction of the companion's light helps and is necessary for placing the Cepheid on its corresponding period-luminosity relation.

The study of the physical parameters starts with the analysis of radial velocity curves, which is much harder than for normal binary systems, as the movement of the photosphere is superimposed on the orbital motion. The observations are very challenging. Moreover, as all the candidates are located in the Magellanic Clouds, at least a 4-m class telescope is necessary to obtain good quality data.

\begin{figure}
\centering
\includegraphics[width=9cm,clip]{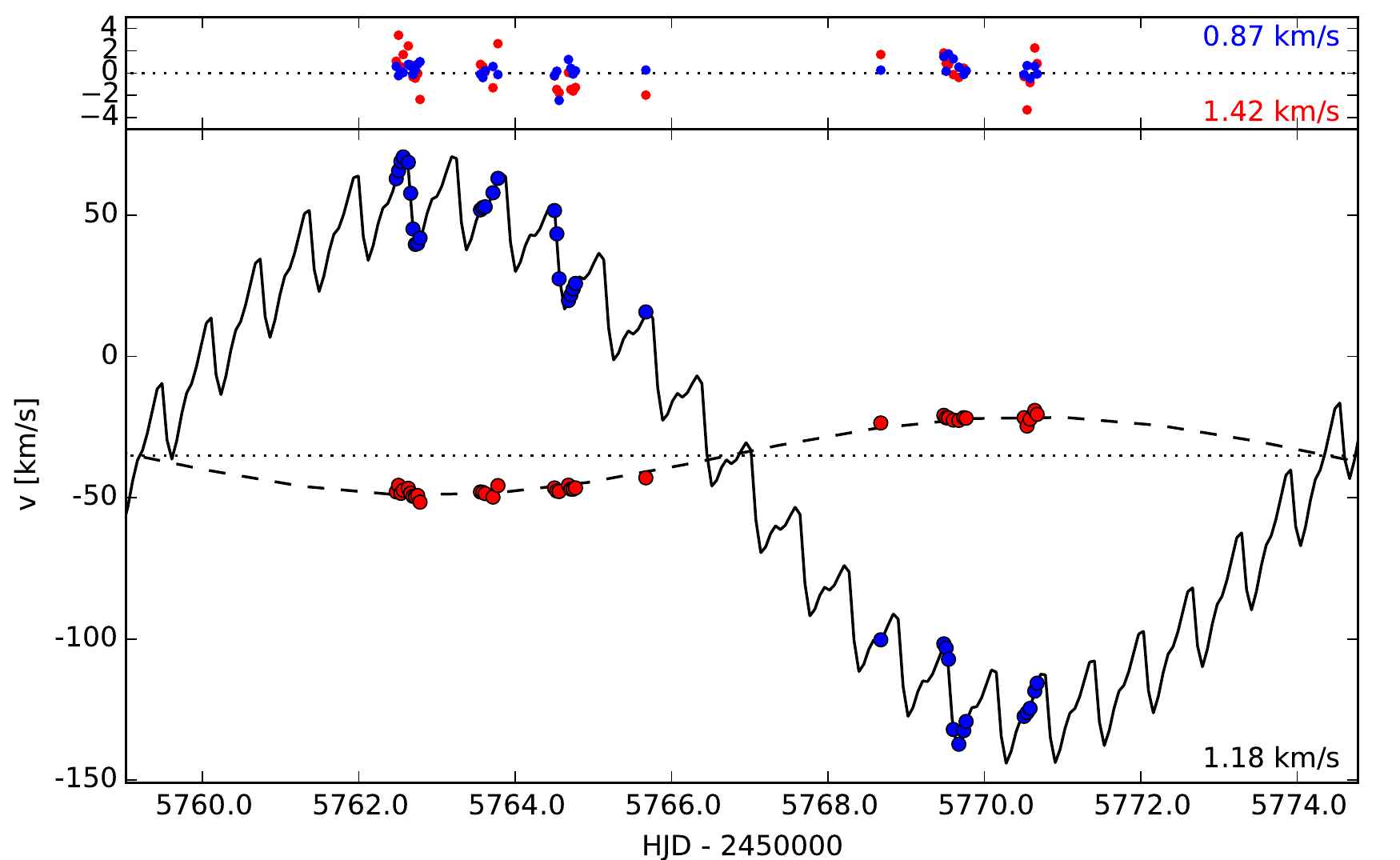}
\caption{Example RV curve with pulsations and the model obtained with the {\tt RaveSpan} code.}
\label{fig:ravespan}
\end{figure}

Once about 20 spectra were collected, the pulsational and orbital motions can be disentangled. We do this using the {\tt RaveSpan} software presented in \citet{t2cep098apj2017}. One of the studied objects is shown in Fig.~\ref{fig:ravespan}. From the analysis of the separated orbital radial velocity curves, we can obtain a value of $m_i\sin^3 i$. For the majority of well-detached systems, it is a good approximation for the masses as $\sin^3 i$ is close to unity. Another important parameter is the system size ($a \sin i$) which is later used to calculate the absolute radii. The pulsational RV curve is used in the modeling of the radius change.

\begin{figure}
\centering
\includegraphics[width=\textwidth]{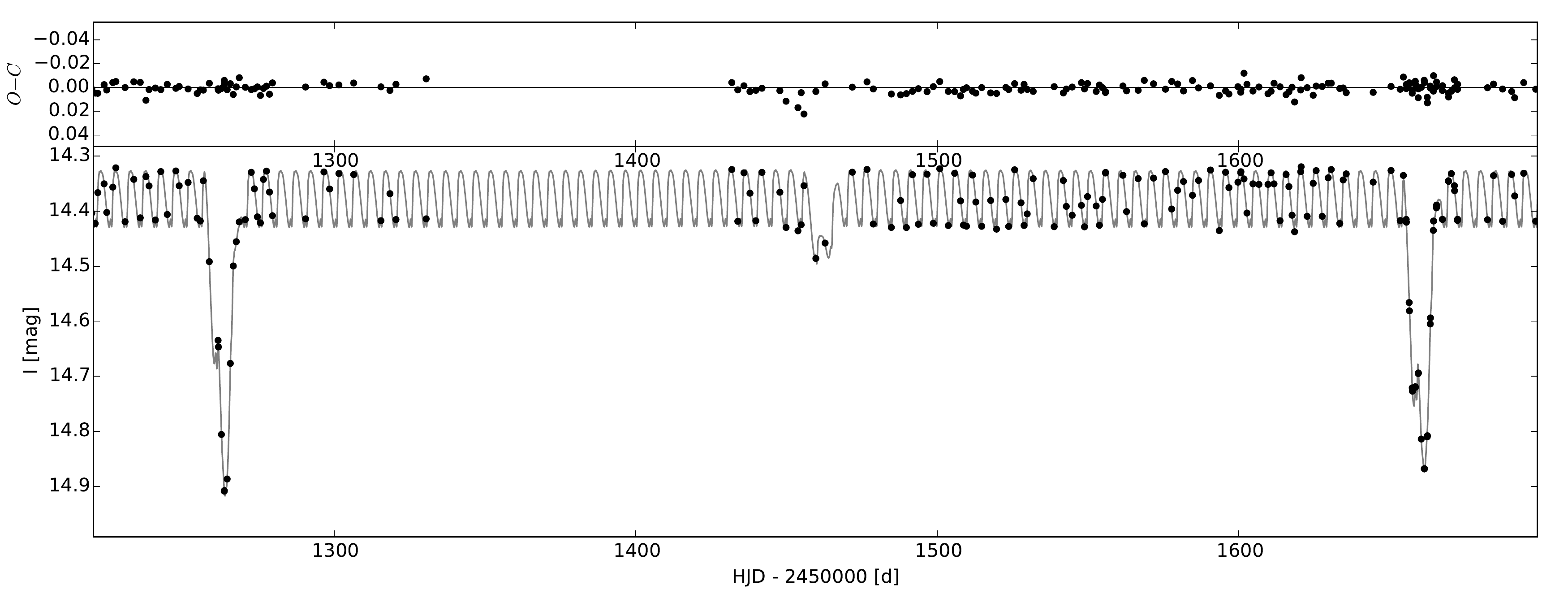}
\caption{$I$-band light curve model of the OGLE LMC-T2CEP-098.}
\label{fig:model098}
\end{figure}

The final modeling uses the orbital solution as a base for the light curve fitting. At this step, the most important information about the physical properties of Cepheids are obtained. The photometric data are analyzed using an eclipsing binary modeling tool based on {\tt JKTEBOP} code \citep{jktebop2004southworth} with the inclusion of pulsation variability. We generate a two-dimensional light curve that consists of purely eclipsing light curves for different pulsating phases. From this two-dimensional model, a one-dimensional light curve is calculated using a combination of pulsational and orbital phases calculated for each observation. As an example, a model for LMC-T2CEP-098 and its $I$-band light curve are shown in Fig.~\ref{fig:model098}. For more technical details on the modeling, please refer to \citet{cep227mnras2013}.

From this solution all the characteristics of the binary system and its components are obtained. For our study, the most important are the {\em radii} of the stars, the exact {\em masses} (we know the inclination at this point), and the individual magnitudes of the components (from the radii and the surface brightness ratio). If observations in more than one filter were analyzed, the colors can be calculated and the temperatures estimated. With these data, the luminosity can be also derived; thus, whenever corresponding data are available, we can obtain all the important physical properties of the Cepheid.

As for now, six systems have been analyzed yielding physical parameters for seven Cepheids. These observational measurements are very important for studying and validating the theoretical pulsational and evolutionary models. For example, our dynamical mass of the LMC-CEP-0227 ($M_{\rm cep} = 4.16 \pm 0.03\,{\rm M}_{\odot}$) showed that the pulsation theory predicts the masses correctly, while the evolution theory gives values about $10\%$ higher \citep{2006MmSAI..77..207B,2008ApJ...677..483K}. Further studies showed, however, that the evolutionary theory also gives consistent results, if the extension of the convective core during the main sequence evolution \citep{Cassisi_Salaris_2011ApJ.728L.43}, mass loss \citep{2011A&A...529L...9N} or rotation \citep{2017EPJWC.15206002A} are taken into account.

\begin{figure}
\centering
\includegraphics[width=8cm,clip]{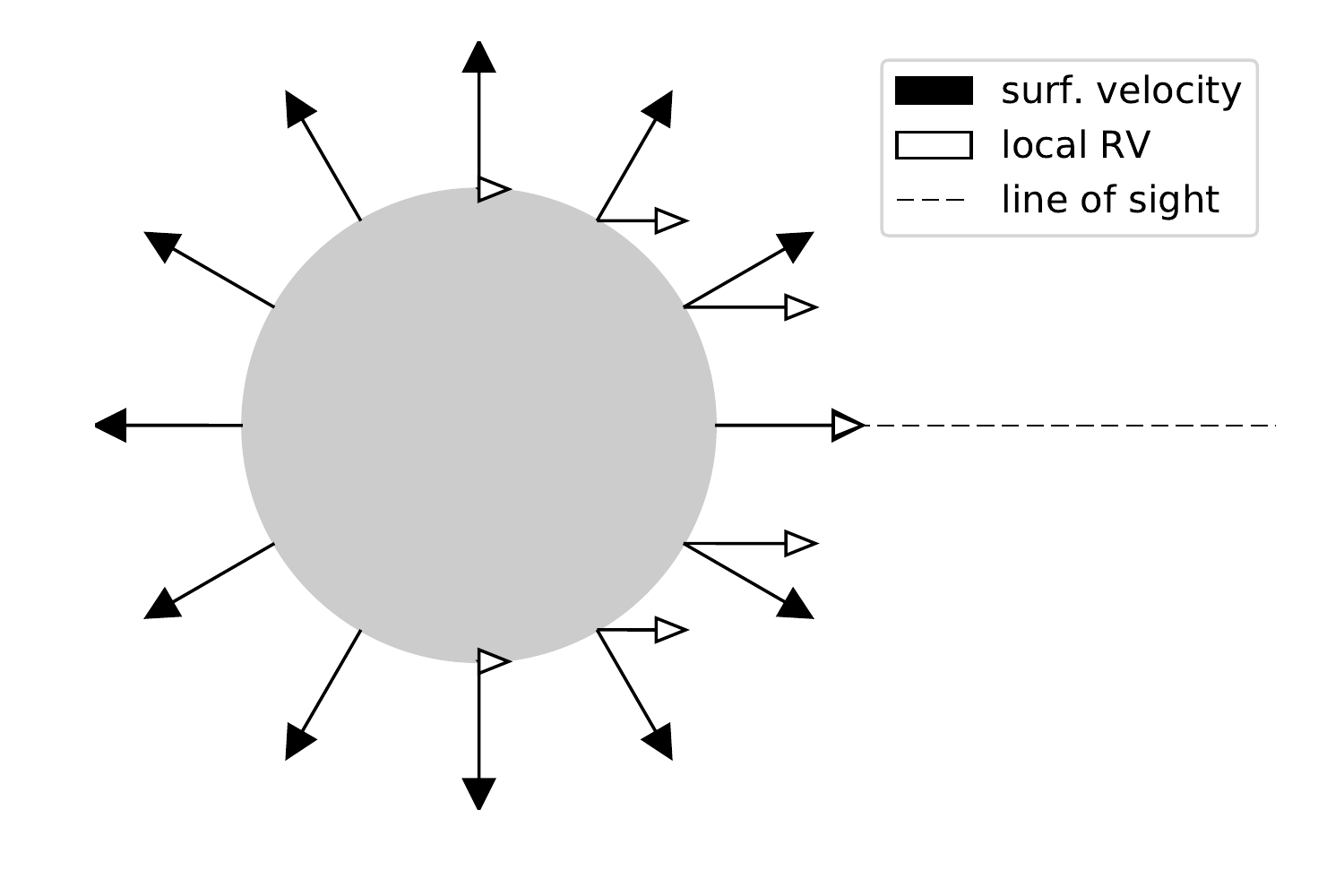}
\caption{The integral over the star disk of the local radial velocities is observed, while for further calculations the surface pulsational velocity is necessary. The $p$-factor is a conversion factor between them.}
\label{fig:pfactor}
\end{figure}

One of the most important parameters we can derive from our models, apart from the masses, is the {\em projection factor} -- a conversion factor between the observed radial velocities and the pulsational surface velocity \citep[e.g.][]{2017A&A...597A..73N}. The general concept is presented in Fig.~\ref{fig:pfactor}. From a practical point of view, it means that the measured velocities should be multiplied by some factor to obtain the true surface velocity.

In our method, the instantaneous radius of the star is a function of the $p$-factor, and the light curve shape depends on it. Depending on the $p$-factor, eclipses in the model may start or end earlier or later, and the amplitude of light variation during the eclipses is also affected. For this reason the quality of our $p$-factor measurements depends strongly on the photometric coverage of the eclipse part of the light curve. Having as many orbital cycles as possible is also beneficial to mitigate the degeneracy between the parameters.

Up to now, we have measured direct, distance-independent $p$-factors for four Cepheids: for LMC-CEP-0227 $p = 1.21 \pm 0.05$, for LMC-CEP-4506 $p = 1.37 \pm 0.09$, and for LMC-T2CEP-098 $p = 1.30 \pm 0.05$. We have also obtained a preliminary $p$-factor for LMC-CEP-1812 \citep{2017EPJWC.15207007P}, yielding $p = 1.26 \pm 0.09$. As the covered period range is short, we cannot exclude the period dependence, but the high scatter of our measurements suggest that other factors may have a bigger impact. For the moment, the anti-correlation with the mass is the clearest one with the higher mass stars having lower $p$-factors. Anti-correlation with the pulsational amplitude is also observed.

\subsection{Distance determination}\label{sec:dist}

\begin{figure}
\centering
\includegraphics[width=9cm,clip]{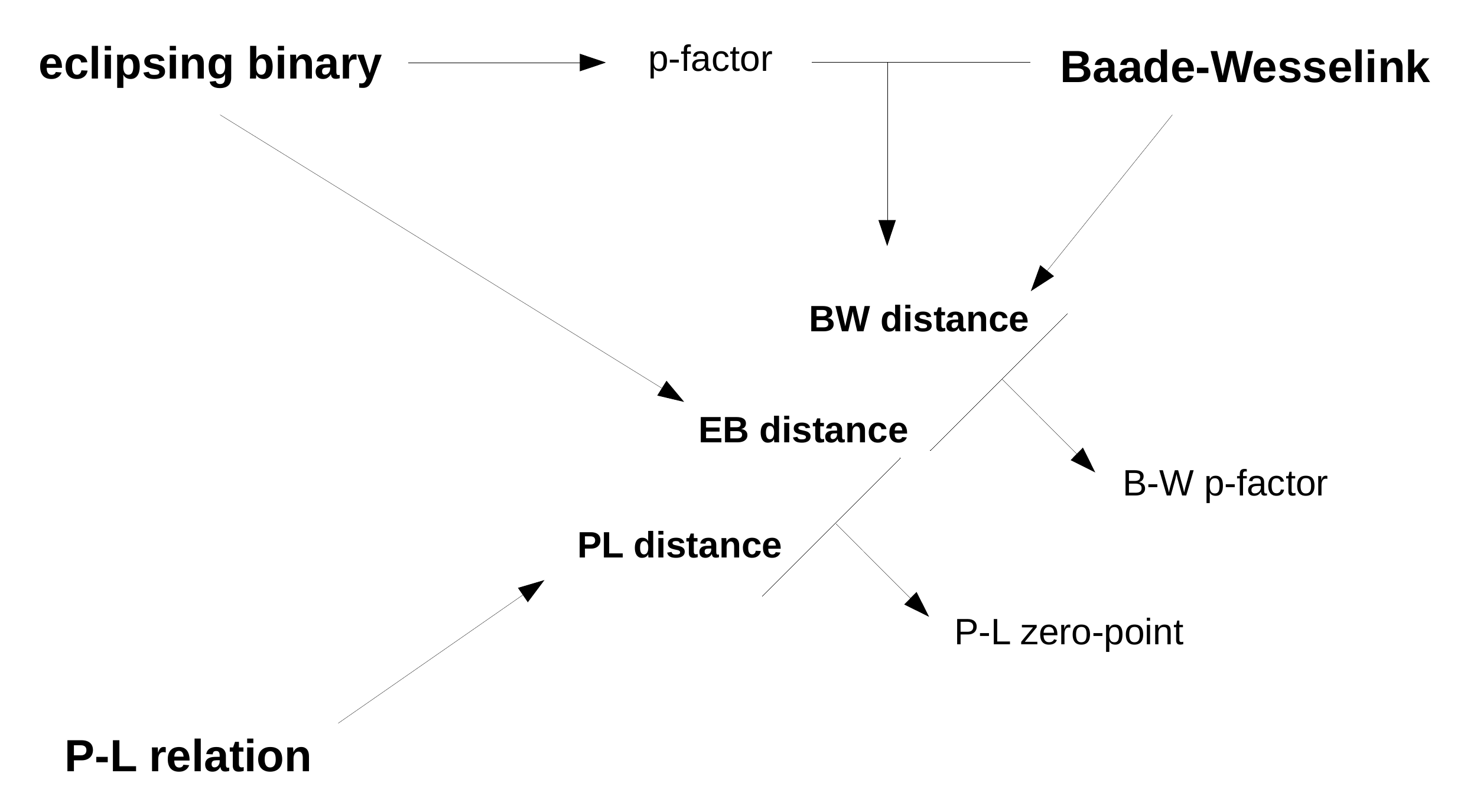}
\caption{A schematic picture of the possible ways to compare and cross-calibrate various distance determination methods to an eclipsing system with a Cepheid.}
\label{fig:distance}
\end{figure}

There are three distinct methods of distance determination we can use for eclipsing binary systems with Cepheid variables: two are based on pulsating stars, and one on eclipsing binary stars. The latter method, being almost purely geometrical, is known for its high precision and accuracy, and presents a great opportunity for comparison with other distance determination methods. The distance from the eclipsing binary method may serve to calibrate the zero-point of the period-luminosity (PL) relation and to estimate the $p$-factor value necessary for the Baade-Wesselink (BW) method. A comparison with the BW method is especially interesting, as we can also use the $p$-factor obtained from the light curve solution. A schematic picture of the calibration possibilities is shown in Fig.~\ref{fig:distance}.

\subsection{Single-lined systems }\label{sec:sb1}

Many candidates for Cepheids in eclipsing binary systems appeared to be single-lined spectroscopic binaries (SB1). This is unfortunate as it gives us only the lower limit for the mass. However, this is true for normal eclipsing systems with stable components. Because of a quite tight relation between the period, mass, and radius for pulsating stars, once we have the period and some constraints for the mass and radius, we can estimate quite reliably all the physical parameters of the pulsating star.

We have done this kind of analysis for the LMC-T2CEP-098, a supposed type II Cepheid in a binary system, for which only the velocities of the pulsating star could be measured. After the subtraction of the pulsational variability, we obtained a relation between the mass of the Cepheid and the size of the orbit. Using the light curve model, this relation can be translated into possible pairs of masses and radii of the Cepheid. For every such pair, we can calculate the period using the pulsation theory and check for which parameters it matches the observed one. The whole procedure is described in \citet{t2cep098apj2017}.

\acknowledgements{We gratefully acknowledge financial support for this work from the Polish National Science Center grant SONATA 2014/15/D/ST9/02248.}

\bibliographystyle{ptapap}
\bibliography{rrlyr2017_konf}

\begin{thebibliography}{18}
\providecommand{\natexlab}[1]{#1}
\providecommand{\url}[1]{\texttt{#1}}
\providecommand{\urlprefix}{URL }
\providecommand{\eprint}[2][]{\url{#2}}

\bibitem[{{Alcock} et~al.(2002)}]{macho2002alcock}
{Alcock}, C., et~al., \emph{{The MACHO Project Large Magellanic Cloud Variable
  Star Inventory. XII. Three Cepheid Variables in Eclipsing Binaries}},
  \emph{\apj} \textbf{573}, 338 (2002), \eprint{astro-ph/0201481}

\bibitem[{{Anderson} et~al.(2017)}]{2017EPJWC.15206002A}
{Anderson}, R.~I., et~al., \emph{{How rotation affects masses and ages of
  classical Cepheids}}, in European Physical Journal Web of Conferences,
  \emph{European Physical Journal Web of Conferences}, volume 152, 06002
  (2017), \eprint{1703.01338}

\bibitem[{{Bono} et~al.(2006){Bono}, {Caputo}, \&
  {Castellani}}]{2006MmSAI..77..207B}
{Bono}, G., {Caputo}, F., {Castellani}, V., \emph{{Stellar pulsation and
  evolution: a stepping-stone to match reality.}}, \emph{\memsai} \textbf{77},
  207 (2006)

\bibitem[{{Cassisi} \& {Salaris}(2011)}]{Cassisi_Salaris_2011ApJ.728L.43}
{Cassisi}, S., {Salaris}, M., \emph{{A Classical Cepheid in a Large Magellanic
  Cloud Eclipsing Binary: Evidence Of Shortcomings in Current Stellar
  Evolutionary Models?}}, \emph{\apjl} \textbf{728}, L43 (2011),
  \eprint{1101.0394}

\bibitem[{{Evans} et~al.(2015)}]{Evans_2015_Cepheid_Binarity_AJ.150.13E}
{Evans}, N.~R., et~al., \emph{{Binary Properties from Cepheid Radial Velocities
  (CRaV)}}, \emph{\aj} \textbf{150}, 13 (2015), \eprint{1505.05823}

\bibitem[{{Gieren} et~al.(2014)}]{cep1718apj2014}
{Gieren}, W., et~al., \emph{{The Araucaria Project. OGLE-LMC-CEP-1718: An
  Exotic Eclipsing Binary System Composed of Two Classical Overtone Cepheids in
  a 413 Day Orbit}}, \emph{\apj} \textbf{786}, 80 (2014), \eprint{1403.3617}

\bibitem[{{Gieren} et~al.(2015)}]{cep9009apj2015}
{Gieren}, W., et~al., \emph{{The Araucaria Project: A Study of the Classical
  Cepheid in the Eclipsing Binary System OGLE LMC562.05.9009 in the Large
  Magellanic Cloud}}, \emph{\apj} \textbf{815}, 28 (2015), \eprint{1511.02826}

\bibitem[{{Keller}(2008)}]{2008ApJ...677..483K}
{Keller}, S.~C., \emph{{Cepheid Mass loss and the Pulsation-Evolutionary Mass
  Discrepancy}}, \emph{\apj} \textbf{677}, 483 (2008), \eprint{0801.1342}

\bibitem[{{Nardetto} et~al.(2017)}]{2017A&A...597A..73N}
{Nardetto}, N., et~al., \emph{{HARPS-N high spectral resolution observations of
  Cepheids I. The Baade-Wesselink projection factor of {$\delta$} Cep
  revisited}}, \emph{\aap} \textbf{597}, A73 (2017), \eprint{1701.01589}

\bibitem[{{Neilson} et~al.(2011){Neilson}, {Cantiello}, \&
  {Langer}}]{2011A&A...529L...9N}
{Neilson}, H.~R., {Cantiello}, M., {Langer}, N., \emph{{The Cepheid mass
  discrepancy and pulsation-driven mass loss}}, \emph{\aap} \textbf{529}, L9
  (2011), \eprint{1104.1638}

\bibitem[{{Pietrzy{\'n}ski} et~al.(2010)}]{cep227nature2010}
{Pietrzy{\'n}ski}, G., et~al., \emph{{The dynamical mass of a classical Cepheid
  variable star in an eclipsing binary system}}, \emph{\nat} \textbf{468}, 542
  (2010)

\bibitem[{{Pietrzy{\'n}ski} et~al.(2011)}]{cep1812apj2011}
{Pietrzy{\'n}ski}, G., et~al., \emph{{The Araucaria Project: Accurate
  Determination of the Dynamical Mass of the Classical Cepheid in the Eclipsing
  System OGLE-LMC-CEP-1812}}, \emph{\apjl} \textbf{742}, L20 (2011),
  \eprint{1109.5414}

\bibitem[{{Pilecki} et~al.(2017{\natexlab{a}}){Pilecki}, {Gieren},
  {Pietrzy{\'n}ski}, \& {Smolec}}]{2017EPJWC.15207007P}
{Pilecki}, B., {Gieren}, W., {Pietrzy{\'n}ski}, G., {Smolec}, R.,
  \emph{{Araucaria Project: Pulsating stars in binary systems and as distance
  indicators}}, in European Physical Journal Web of Conferences, \emph{European
  Physical Journal Web of Conferences}, volume 152, 07007 (2017{\natexlab{a}})

\bibitem[{{Pilecki} et~al.(2013)}]{cep227mnras2013}
{Pilecki}, B., et~al., \emph{{Physical parameters and the projection factor of
  the classical Cepheid in the binary system OGLE-LMC-CEP-0227}}, \emph{\mnras}
  \textbf{436}, 953 (2013), \eprint{1308.5023}

\bibitem[{{Pilecki} et~al.(2015)}]{cep2532apj2015}
{Pilecki}, B., et~al., \emph{{The Araucaria Project: the First-overtone
  Classical Cepheid in the Eclipsing System OGLE-LMC-CEP-2532}}, \emph{\apj}
  \textbf{806}, 29 (2015), \eprint{1504.04611}

\bibitem[{{Pilecki} et~al.(2017{\natexlab{b}})}]{t2cep098apj2017}
{Pilecki}, B., et~al., \emph{{Mass and p-factor of the Type II Cepheid
  OGLE-LMC-T2CEP-098 in a Binary System}}, \emph{\apj} \textbf{842}, 110
  (2017{\natexlab{b}}), \eprint{1704.07782}

\bibitem[{{Southworth} et~al.(2004){Southworth}, {Maxted}, \&
  {Smalley}}]{jktebop2004southworth}
{Southworth}, J., {Maxted}, P.~F.~L., {Smalley}, B., \emph{{Eclipsing binaries
  in open clusters - II. V453 Cyg in NGC 6871}}, \emph{\mnras} \textbf{351},
  1277 (2004), \eprint{astro-ph/0403572}

\bibitem[{{Udalski} et~al.(2015){Udalski}, {Szyma{\'n}ski}, \&
  {Szyma{\'n}ski}}]{ogle2015udalski}
{Udalski}, A., {Szyma{\'n}ski}, M.~K., {Szyma{\'n}ski}, G., \emph{{OGLE-IV:
  Fourth Phase of the Optical Gravitational Lensing Experiment}}, \emph{\actaa}
  \textbf{65}, 1 (2015), \eprint{1504.05966}

\end{thebibliography}

\end{document}